\documentclass[jcs]{iosart2x}

\pubyear{0000}
\volume{0}
\firstpage{1}
\lastpage{1}

\usepackage{color}
\usepackage{caption}
\usepackage{amsmath}
\usepackage{relsize}
 
\def\skipnoindent{\vskip0.1in\noindent}

\def\state#1{{\tt #1}}

\newcounter{exnum}

\usepackage{amssymb}
\usepackage{graphicx}

\usepackage{url}

\usepackage{amsmath}
\usepackage{array}

\usepackage{xspace}
\def\state#1{{\tt #1}\xspace}

\begin{document}
\begin{frontmatter}
   
\title{
Investigation of Cyber Attacks on a Water Distribution System }
\runningtitle{Investigation of cyber attacks ICS: WADI}

\author[A]{\inits{N.}\fnms{Sridhar} \snm{Adepu}\ead[label=e1]{adepu\_sridhar@mymail.sutd.edu.sg}%
\thanks{Corresponding author. \printead{e1}.}},
\author[A]{\inits{N.}\fnms{Venkata Reddy} \snm{Palleti}\ead[label=e2]{venkata\_palleti@sutd.edu.sg}},
\author[A]{\inits{N.}\fnms{Gyanendra} \snm{Mishra}\ead[label=e3]{f2013126@pilani.bits-pilani.ac.in}}
and
\author[A]{\inits{N.}\fnms{Aditya} \snm{Mathur}\ead[label=e4]{aditya\_mathur@sutd.edu.sg}}
\runningauthor{Adepu et al.}
\address[A]{iTrust Center for Research in Cyber Security, \orgname{Singapore University of Technology and Design},
\printead[presep={\\}]{e1,e2,e3,e4}}



\newcommand{\placetextbox}[3]{
\setbox0=\hbox{#3}
\AddToShipoutPictureFG*{ \put(\LenToUnit{#1\paperwidth},\LenToUnit{#2\paperheight}){\vtop{{\null}\makebox[0pt][c]{#3}}}
}
}

\thispagestyle{empty}

\begin{abstract}

A Cyber Physical System (CPS) consists of cyber components for computation and communication, and physical components such as sensors and actuators for process control. These components  are networked and interact in a feedback loop. CPS are found in critical infrastructure such as water distribution, power grid, and mass transportation. Often these systems are vulnerable to attacks as the cyber components such as Supervisory Control and Data Acquisition  workstations, Human Machine Interface  and Programmable Logic Controllers  are potential targets for attackers.  In this work, we report a study to investigate the impact of  cyber attacks  on a water distribution (WADI) system. Attacks were designed to meet attacker objectives and  launched on WADI using a specially designed tool. This tool enables  the launch of single and  multi-point attacks where the latter are designed to specifically hide one or more attacks. The outcome of the experiments led to a better understanding of attack propagation and behavior of WADI in response to the attacks as well as to the design of an attack detection mechanism for water distribution system.  
 
\end{abstract}
  
\end{frontmatter}


\section{Introduction}
\label{sec:introduction}
 
\begin{figure}[htbp]
\centering
\includegraphics[scale=0.8]{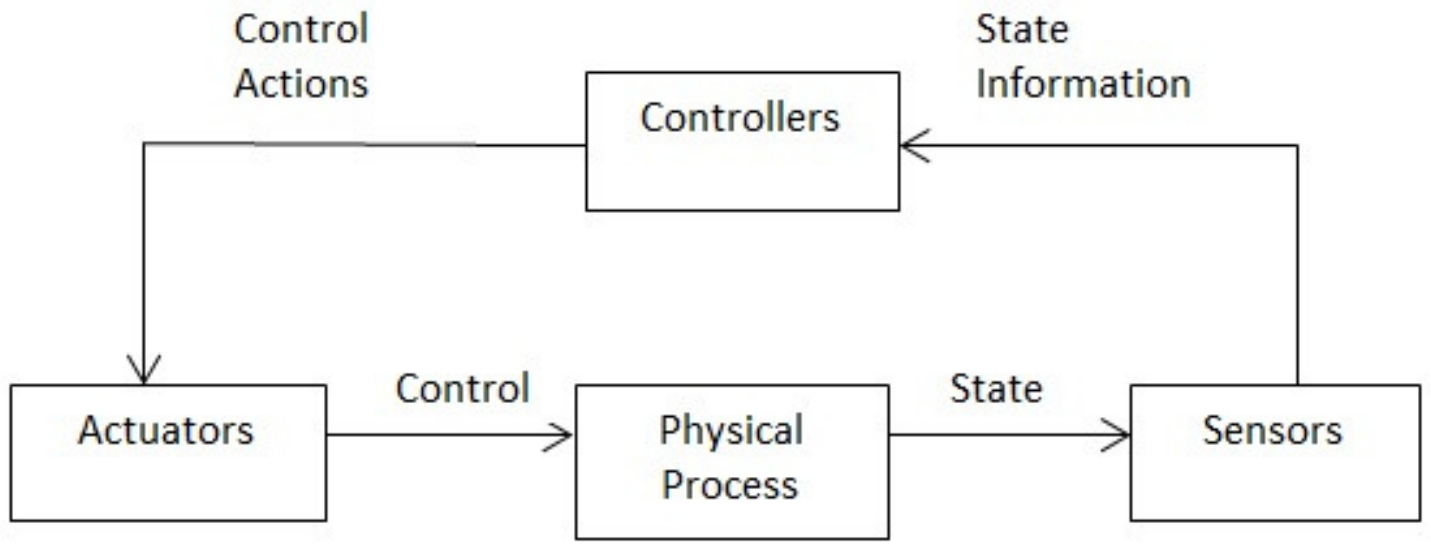}
\caption{\small Key components in a CPS. State transformation of a CPS in a feedback control loop. \normalsize}
\label{fig:stateTransformer}
\end{figure}

Cyber Physical Systems (CPSs) are found in critical infrastructure such as water distribution, energy and transportation. CPS consists of a physical process controlled by an Industrial Control System (ICS). In a CPS, a set of sensors measure process variables such as temperature, flow rate, level etc., from the physical process and send these values to the controllers through communication channels. Based on these values the controller makes decisions and initiates actions on the physical process. Figure\,\ref{fig:stateTransformer} shows the representation of a CPS as a feedback  system\,\cite{adepuMathurCSDM2016}.     

Attacks on ICS can have a significant impact depending on the type of attack and its location. The increase in successful cyber attacks on ICS\,\cite{ukraineBlackout,weinbergerStuxnet}, and many  unsuccessful attempts\,\cite{ics-cert}, points to the importance of research in the design of ICS that is resilient to cyber attacks. Attacks are a result of exploitation of one or more vulnerabilities in an ICS. Such vulnerabilities  might be  due to the lack of access control in the system\,\cite{adepuMishraMathurQRS2016}, software vulnerabilities in the Programmable Logic Controllers (PLCs), Supervisory Control and Data Acquisition (SCADA) software systems, and weaknesses in the communication channels. 

\skipnoindent{\it Motivation:} Several attacks on water distribution systems have been reported  in recent years such as the  Kemuri Water Company (KWC)\footnote{http://www.securityweek.com/attackers-alter-water-treatment-systems-utility-hack-report} attack,  in 2016. The attack  resulted in the exposure of personal information of the utility's 2.5\,million customers. Reports from ICS-CERT\,\cite{icsCERTAdvisory} indicate that an understanding of these attacks against critical infrastructure is important for rapid  investigation and evaluation of detection methods.  The work presented in this paper is a  step towards realizing a safe and secure  water distribution infrastructure. 
To create  effective protection methods that lead to low false alarm and high detection rates, one needs to understand the nature of  attacks on water distribution systems and the system response.

\skipnoindent {\em Goals and research questions}: The goal of the study reported here is to (a)\,understand vulnerabilities and design potential attacks and (b)\,investigate the impact of cyber physical attacks. The following questions are addressed through experimentation on WADI: 
 \noindent \textbf{RQ1:} {\em How do cyber attacks impact a water distribution system?}
 \noindent \textbf{RQ2:} {\em How does knowledge of the response of a water distribution system to one or more cyber attacks help in designing an attack detection mechanism?}
 
\skipnoindent{\em Contributions}: In the context of a specific water distribution plant: (a)\,A tool to launch attacks and (b)\,design and implementation of attacks on a water distribution system.
 
\skipnoindent{\em Organization}: The remainder of this paper is structured as follows. Background and preliminary works are explained in Section\,\ref{sec:background}. 
Section\,\ref{sec:context-wadi} presents the context of this work and includes  architecture of WADI, vulnerability assessment, and how attacks can be launched on WADI. Section\,\ref{sec:casestudy} describes the attack design and investigation on WADI. 
Response to the research questions and lessons learned are discussed in Section\,\ref{sec:discussion}. Related work is presented in Section~\ref{sec:relatedwork}. Section\,\ref{sec:conclusions} offers a summary of this work and future work.

\section{Preliminaries and background}
\label{sec:background}

\begin{figure}[ht]
\centering
\includegraphics[scale=0.6]{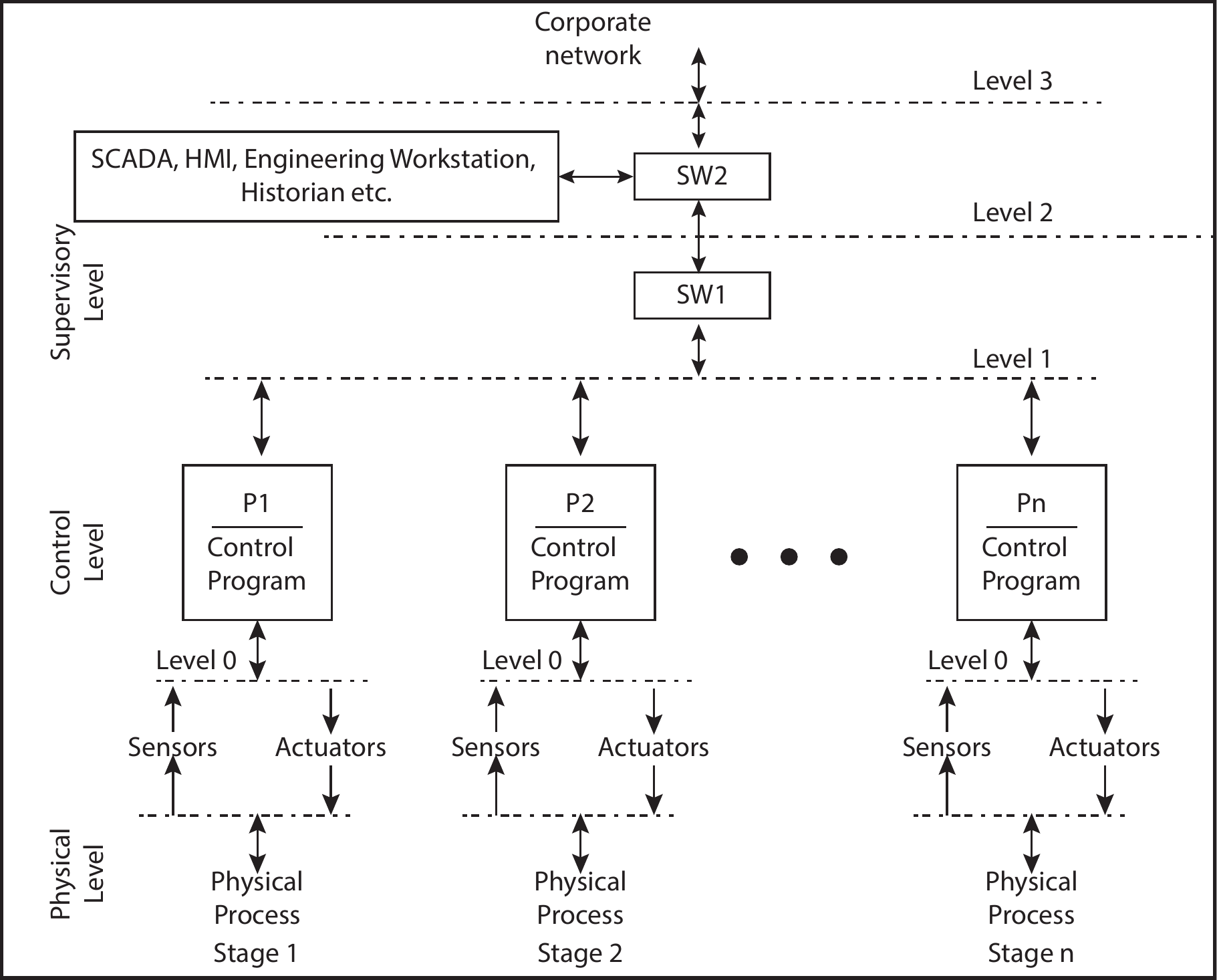}
\caption{\small Architecture of the control portion of a CPS. P1, P2,\ldots,Pn denote PLCs. Each PLC communicates with its sensors and actuators through a local network at Level~0. PLCs communicate among themselves via another network at Level~1.  Communication with SCADA and other computers is   not shown here. \normalsize}
\label{fig:CPSStructure}
\end{figure}


This section provides information needed to understand the remaining paper. 

\subsection{Industrial Control Systems}
\label{sec:ICS}
ICSs are found in plants such as  water treatment,  distribution, and in power generation, transmission and distribution. The complexity of an ICS increases the attack surface for an attacker to launch attacks both at the cyber-and the physical-parts of a plant. Control software in an ICS may also contain vulnerabilities for reasons such as un-patched or practically impossible to patch legacy code, the absence of standard security certifications for ICS devices, and the lack of resources to keep the ICS updated. 

\skipnoindent{\em Communication Structure of ICS:} ICS consist of distributed supervisory control systems. The control system itself is a collection of PLCs, each controlling a specific portion of the physical process. Each PLC communicates with a set of sensors and actuators via a local network (Figure\,\ref{fig:CPSStructure}) through a multi-layer network also referred to as the {field-bus} network\,\cite{stoufferFalcoScarfone}. The PLCs communicate with each other using the Level~1 network. Such a layered network is in accordance with the prevailing practice for ICS\,\cite{gallowayHancke}.
As mentioned in Section\,\ref{sec:introduction}, attacks on ICS are on the rise. The results of a recent survey\,\cite{threatLandscape} show on threat landscape on ICS in September~2017. It represents the attack space and how often an attacker attempts to enter an ICS. Such attempts, often successful, motivate the study reported here.

SCADA and Distributed Control Systems are referred as to as operational Technology (OT). The convergence of Information Technology (IT) and OT\,\cite{MurrayJohnstoneValli} is increasing in water distribution systems.  With this convergence, OT data is now accessible from IT environment such as via remote access. The OT data includes critical information regarding the plant such as temperatures, level indicators, control signals, sensor signals and actuator statuses; especially  so in water distribution systems as they are distributed across a  city making it an easy target for cyber-physical  attacks.

\subsection{Vulnerability Assessment}
\label{sec:vulnerability-assessment}

Vulnerability assessment on ICSs follows  four main steps\footnote{https://www.secureworks.com/blog/vulnerability-assessments-versus-penetration-tests}: 1)~identify list of assets and resources in the system, 2)~assign importance to the resources, 3)~identify security vulnerabilities in each asset and resource, 4)~propose mitigation for the most serious vulnerabilities. 

In order to know all the vulnerabilities in ICS, one must know the associated paths within ICS communications. In\,\cite{DHS_vulnerabilities_ICS} authors explained different paths through which an attacker can enter into the system using various devices, communications paths, and methods that can be used for communicating with process system components. 
An attacker who wishes to attack ICS has to go through the following steps: 1)~gain access to the ICS network 2)~perform reconnaissance and understanding of the process 3)~gain control of ICSs.

\skipnoindent Some of the industries conducted the vulnerability assessments in industrial systems and published the results. Following are the summary of reports from Kasper-sky and Honywell. 
\noindent {\em Kasper-sky}\,\cite{KL_REPORT_ICS_Statistic_vulnerabilities} summarized the findings of it's research on ICS vulnerabilities as follows:  
Over the years, 19 vulnerabilities in 2010 increased to 189 vulnerabilities in 2015. Even though the vulnerabilities are fixed by the product manufactures, the ICS management not upgrading soon. At least 5\% of the vulnerabilities published by ICS-CERT were not fully fixed. Sometimes the vulnerable component was removed from the market and vendor support may not be available anymore.
\noindent {\em Honeywell XL Web II Controller Vulnerabilities}\,\cite{vulnerability-ICSA-17-033-01} are found by an independent researcher. An attacker may use these to expose a password by accessing a specific URL. The XL Web II becomes an entry point into the network.

 \begin{figure*}[htbp]
\centering
\includegraphics[scale=0.45]{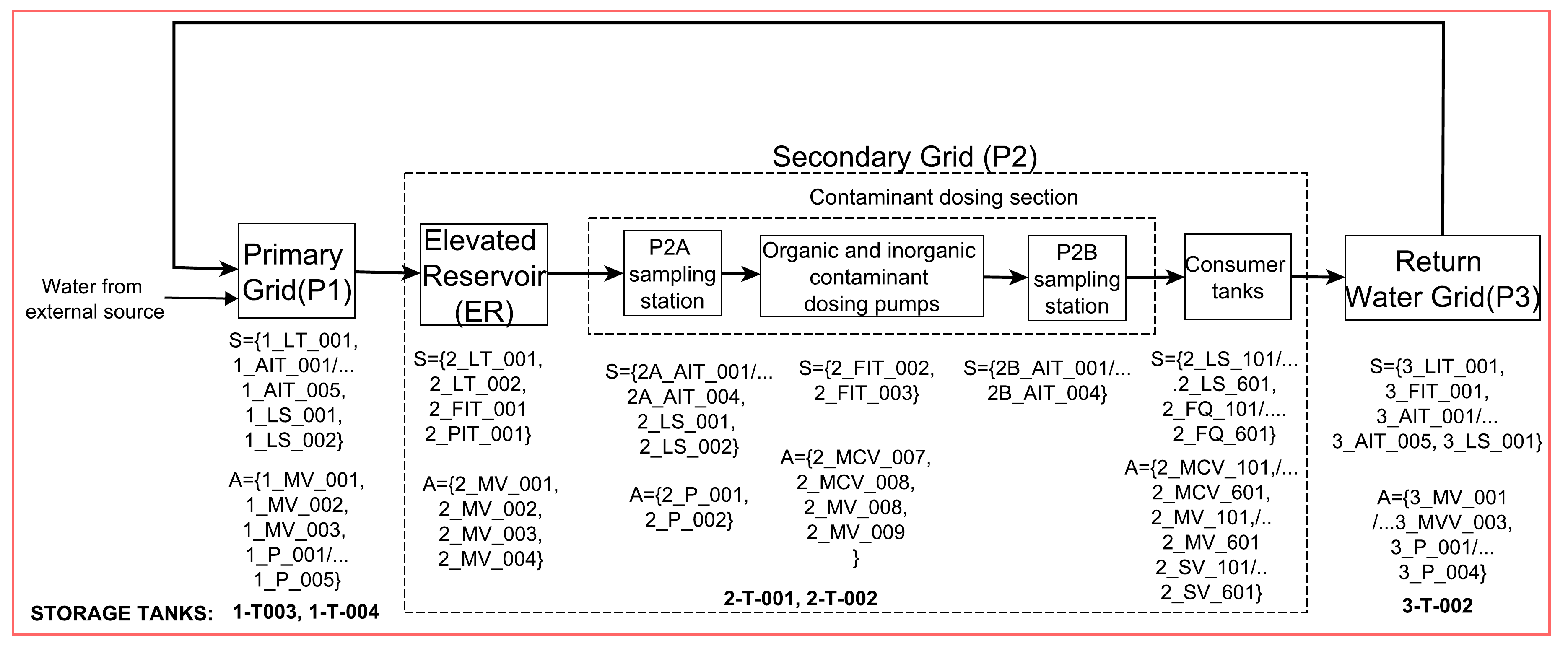}

\caption{ \small Three stages in WADI are shown. Solid arrows indicate flow of water and sequence of processes. S: set of sensors;  A: set of actuators.  LT-Level Transmitter, AIT-Analyzer Indication Transmitter, FIT-Flow Indication Transmitter, PIT-Pressure Indication Transmitter, LS-Level Switch. Actuators: P-Pump, MV-Motorized valve, MCV-Modulating Control Valve, SV-Solenoid Valve. Tag name of the instrument is indicated as XXX\_YYY\_ZZZ, where XXX, YYY and ZZZ  represent stage number, instrument type and instrument index, respectively.   \normalsize } 
\label{fig:wadi}

\end{figure*}

\section{Context: WADI Testbed}
\label{sec:context-wadi}
 This study centers around a Water Distribution (WADI) testbed\footnote{\url{https://itrust.sutd.edu.sg/research/testbeds/water-distribution-wadi/}}.  This section covers the testbed architecture and the communication channels.

\subsection{Architecture of the WADI}
\label{sec:wadi}

Water distribution (WADI) plant\,\cite{mujeebPalletiMathur} is an operational testbed supplying 10\,US gallons/min of filtered water. It represents a scaled-down version of a large water distribution network in a city. WADI consists of three stages (Figure\,\ref{fig:wadi}), namely primary grid (P1), secondary grid (P2), and return water grid (P3). Primary grid consists of two raw water (RW) tanks of 2500 liters each. These tanks are fed by three incoming sources including  Public Utility Board (PUB), return water grid, and from a water treatment plant. A level sensor (1\_LT\_001) is installed in the primary grid to monitor the levels in the RW tanks. Water quality analyzers are installed to measure pH, turbidity, conductivity and residual chlorine. Secondary grid consists of two Elevated Reservoir (ER) tanks, consumer tanks, and contamination sampling stations. RW tanks supply water to the ER tanks using raw water pump (1\_P\_003) which is installed in the primary grid. Two level sensors, 2\_LT\_001 and 2\_LT\_002 are installed in ER tanks to measure water levels. Further, water from ER tanks flows into the consumer tank based on the preset demand pattern. 

Two water quality monitoring stations are installed at consumer tanks. One  station is at  the immediate downstream of reservoir and another is before the consumer tanks (P2A and P2B stations in Figure\,\ref{fig:wadi}). These stations ensure water quality before it is sent to the consumer tanks. Once a consumer tank is filled, a level switch installed  raises an alarm and water from the tank drains into the return water grid. To recycle  water, return water grid pumps water to the primary grid.  Water quality analyzers are installed in return water grid to check water quality before pumping it into the primary grid. 

Three PLCs are installed to control each stage of WADI. These PLCs use CompactRIO as RIO (Remote Input Output)  from National Instruments. In addition to the PLC in the secondary grid, two Schneider Electric Remote Terminal Units (RTUs), which use SCADAPack, are installed to measure water quality.  There is a total of  103 sensors and actuators operating to measure  water levels, water quality, flow rates, pressure, and  status of motorized valves and pumps. There are three levels of networks in WADI. Level~0 corresponds to the communication between PLC's and sensors  over Modbus RS485. Level\,1 corresponds to communications using  the National Instrument's publish subscribe protocol (NI-PSP) while the SCADAPack RTUs communicate through Modbus TCP. PLCs at  Stage-1 and Stage-3 are connected to analyzers capable of communicating through Modbus Serial. Level\,2 consists of communication between the HMI and the plant control network. The interconnection of HMI, workstations and PLCs allows remote monitoring.

\subsection{Vulnerability Assessment in WADI}
\label{assessment-WADI}


To identify  vulnerabilities in an ICS, one must know the associated paths within its communication infrastructure.  In\,\cite{DHS_vulnerabilities_ICS} authors explained paths through which an attacker can enter the system using various devices, communications paths, and methods that can be used for communicating with process system components.



\skipnoindent{\em List of assets and resources in the system:} The list of assets are mentioned in the Table~\ref{assets}. In this subsection, different vulnerabilities in the WADI are explained based on the assets provided from  Table\,\ref{assets}. These includes Eternal Blue, default admin password on web server, and vulnerabilities in restful web service, Modbus serial and TCP, objective C program that speaks NI-PSP and custom VI that interacts with a python script.

\begin{table*}[tbh]
\centering
\caption{Assets Table}
\vspace{0.5 cm}
\label{assets}
\begin{tabular}{|p{1 in}|p{1.5 in}|p{2.8 in}|}
\hline
{\bf Asset} & {\bf Version/Model used} & {\bf Location} \\ \hline
SCADA System & SCADA System from Labview is used for the application.  & SCADA System computer running on Windows 7. \\ \hline 

PLCs & NI PLC is used in WADI to control various operations & Control and network panel and works based on the firmware and control logic program. Communicates with NI-PSP and Modbus TCP/IP communication in few cases \\ \hline

Network Switches & Moxa ES5 301 & Network Control panel \\ \hline

Access points &  Wifi access points & Network Control panel \\ \hline
\end{tabular}
\end{table*}

\skipnoindent {\em Eternal Blue:}\,\cite{TristanChristosDavid,kharraz2017techniques} This is an exploit that focuses on Microsoft Windows and used for the wannacry ransomware attack in 2017. EternalBlue\,\cite{nakashima2017nsa} is vulnerability in server message block (SMB) protocol. This is mentioned in  CVE-2017-0144\,\cite{eternalBlue-CVE} catalog. SMB server mishandles the packets from remote attackers, which eventually allows to access to the system. Attacks similar to wannacry attack was studied in automotive sector\,\cite{zimba2018multi} and identified as an emerging threat to critical infrastructure and industrial control systems. 

\skipnoindent {\em Default Admin Password on webserver:}\cite{icsCERTAdvisory,ICSCERTFY2015,DHS_vulnerabilities_ICS} Manufacturures follow default passwords, and during the installation and configuration period, the operating management are not changing the default passwords. Attacker can use those default passwords from each manufacturing unit and exploit the system. Later it could be used to modify the functions of the overall control the system.

 To develop the attack tool all communication channels were studied and investigated for openings and vulnerabilities. A lot of them lacked any form of access control. Different parts of WADI support various different communication channels like MODBUS between RTUs and SCADA, NI-PSP between various controllers and RTUs.

\skipnoindent{\em Restful Web Service:}
LabVIEW allows VIs to be equipped with restful web services which manipulates the  data via HTTP methods like GET, POST etc. These services don't require any authentication by default.

\skipnoindent{\em Modbus Serial And TCP:} 
RTUs P2A and P2B run Modbus TCP while the analyzers installed in  P1 and P3 are connected via Modbus serial. 
The protocol was designed with safety in mind but not security and hence lacks any type of access control, if you can ping a device running Modbus you can own the device.
Python has a couple of libraries which speak Modbus, most importantly pyModbus. Using this library an attack tool was designed capable of reading and manipulating data on 8 sensors connected in P2A and P2B related to water quality. These sensors are responsible for measuring the water properties such as pH, ORP, conductivity.


\skipnoindent{\em Objective C Program that Speaks NI-PSP:} 
It was found that there exist C\# and Visual Basic libraries that speak NI-PSP. These libraries are proprietary and consist of  Measurement Studio  for Visual Studio. This allows any attacker to write and read basic data from sensors and actuators in the plant.  

\skipnoindent{\em Custom VI that interacts with a python script:} 
This method relies on using special VIs (Virtual Instruments) or LabVIEW programs that can read and write to the cluster variables.  To make this method more dynamic, a python package was written that could speak to the Virtual Instruments to craft more complex attacks giving complete access to the system. The NI-PSP implementation in the water distribution system plant has no authentication or access control as mentioned above. As long as an adversary can access the network they can control the entire plant. 

 \skipnoindent{\em Summary:} All the above methods rely on the fact that the network is very open. The system has no authentication in place and depends on the network to be full of good nodes acting in the interest of the plant.

\begin{itemize}
\item The National Instruments Publish Subscribe Protocol variables have a property through
which they can be made accessible to certain users/groups through an additional plugin but the configuration of the plant
allows access to any user on any host as long as they can connect with the PLC/SCADA
system. 
The publish subscribe protocol has no security for variables by default. One has to pay for another product
called DSC or Datalogging and Supervisory Control to have any form of security.\\

\item Modbus is known to be very open and insecure. As long as one can assume the IP address
of one of the registered devices in the network, one can access, read and write any variable
on any register via Modbus. Assuming IP address is as simple as removing one of the
cables from one of the switches and plugging in your own cable.
Despite the lack of any access control methods, MODBUS finds itself being continously used in a lot of Industrial Control
Systems. It has no passwords, no authorization, no facility to pass certificates but it continues to be used because
of it's popularity and simplicity. Having a firewall in place is one of the methods to ensure that a PLC isn't exposed
to the internet but this doesn't solve the inherent problems that MODBUS brings, it only pushes it up a level. Now the attacker has to access a machine that the firewall trusts in order to gain access to the PLC supporting MODBUS. According to the Internet of Things Search Engine Shodan, there are 17,000 devices listening to MODBUS on the internet majority of them being in the united states. \footnote{
https://blog.shodan.io/content/images/2015/09/screenshot-www-shodan-io-2015-09-04-22-33-29.png}
\end{itemize}

CWE-284\footnote{https://cwe.mitre.org/data/definitions/284.html} Improper Access Control talks about systems with improper or no access control. The state of the plant at the moment exhibits CWE-284. As an adversary can easily read and manipulate critical data in the plant the plant at the moment is also guilty of having the CWE-306: Missing Authentication for Critical Function weakness.

 Moving forward protocol designers and software/PLC component manufacturers should push for proper authentication by default. The lack of secure defaults is no minor issue and is nothing new. A lot of studies have shown the impact of insecure defaults and how users generally don't change\footnote{https://www.uie.com/brainsparks/2011/09/14/do-users-change-their-settings/}\small{'}\footnote{https://www.nngroup.com/articles/the-power-of-defaults/} the defaults if they don't have to.
 


\subsection{Attacking WADI}
\label{attacking}
 As mentioned in Section\,\ref{sec:wadi}, WADI uses  a multi layered network comprising of different protocols at different levels and between different devices. For this paper the focus is on the National Instruments Publish Subscribe Protocol (NI-PSP). NI-PSP is the most used protocol in the entire WADI network and provides access to all data on the network. We developed an attack tool named NiSploit\footnote{https://gitlab.com/gyani/NiSploit} that uses custom LabVIEW Virtual Instruments (VIs) that communicate with  shared variables present on different PLCs across the plant using NI-PSP. Earlier exploration into various other mechanisms gave limited access to the variables\,\cite{adepuMishraMathurQRS2016}. 
 
 Shared variables are used by a controller and SCADA to expose data over the network via a shared variable engine. These variables reside in controllers and the SCADA, have publish-subscribe architecture, and are shared using the NI-PSP. Network shared variables publish data through the shared variable engine. The shared variable engine resides on a SCADA and manages variables using the NI-PSP protocol. In the publish subscribe model the publishers do not publish to clients; instead they send data to the shared variable engine after every update and the subscribers subscribe to the shared variable engine for changes.

LabVIEW programs, or VIs, are simple drag and drop programs. We have written custom VIs for the purpose of attacking the National Instruments Publish Subscribe Protocol Variables. Several different custom VIs have been created, each one for  attacking different types of cluster variables used in WADI. The Python module is the front end of the tool and an attacker needs to be concerned only with the use of this module. The module uses ActiveX\,\cite{activeX} to control the LabVIEW application from python code. It connects to ActiveX controls using the Pywin32 library. ActiveX allows the user to run programs and specific functions that the program has exposed via it's ActiveX server. LabVIEW exposes a lot of different functionality including the ability to run VIs, set values for different controls and to fetch values of interest. The custom VIs along with the python module allow for creating powerful and complex controlled attacks. The attacks designed and executed in the following Section (Section\,\ref{sec:casestudy}) are realized through the NI-PSP attack tool called NiSploit.



\section{Attack Investigation on WADI }
\label{sec:casestudy}

This section presents a detailed case study which includes attack design, execution of attacks and results. We assumed an attacker\,\cite{Rocchetto2016} has an ability to enter into the system through vulnerabilities and social engineering. Further, we considered an insider attacker profile in which attacker has the process, communication knowledge, and access to the communication channels.   

\subsection{Attack Design}
\label{sec:attackDesign}

Attacks considered in this paper are launched on primary grid (P1) and secondary grid (P2) of WADI (Section~\ref{sec:wadi}). 
Stage-1 contains a tank whose level is measured by 1\_LT\_001. The stage-2 tank is responsible for water received by the consumer and its level is measured by 2\_LT\_002. Valve 1\_MV\_001 is responsible for the flow of water from RW tanks to the drain. Valve 1\_MV\_002 is responsible for the inflow of water to the RW tank. Valve  2\_MV\_003 is responsible for inflow of water to the ER tank. Water flows from the RW tank to the ER tank. In this study, an attacker is an insider, who has an access to the system: process, communication knowledge, and access to the communication channels. 



  Cyber attacks on WADI were derived from a CPS-specific generalized attacker model\cite{adepuMathurCompsac2016,adepuMathurHASE2016}. This  model contains the attacker's intents (set $I$), and the attack domain ($D$). For example, in a water distribution system attacker's intent could be water pump damage or overflow the water from a tank. An attack model  for a CPS is represented as a six-tuple $(M, G, D, P, S_o , S_e )$. An attack procedure $M$ is designed by the attacker to realize an attack on a finite set of attack points $P$ in a CPS when this CPS is in state $S_o$, and possibly removed when the CPS is in state $S_e$. This attacker model is useful in generating a variety of attacks. Attack procedure $M$ contains the attack vectors which include how an attacker enters into the system and manipulate different communication channels. The procedure $M$ essentially  the use of the NiSploit tool as described in Section~\ref{attacking}. Goal $G$ is equal to Intent $I$. Domain $D$ is derived from the CPS domain\,\cite{adepuMathurCompsac2016}. For each CPS, domain is different based on the kind of physical process and components involved. Here, $P$ is a set of  sensors, actuators or any other potential attack points.  $S_o$ is the starting state of the system at the time of attack launch starting and $S_e$ is the end state of the system when the attacker ends an attack. When $S_e$ and $I$ is identical then it shows that attacker reached his intent or attacker made an impact on the system.  
 
 Impact of  attacks can be viewed along three\,\cite{adepuMathurCompsac2016} dimensions: $(C_m,P_r,P_e)$, where $C_m$ represents the impact on components of the system, $P_r$ is the impact on properties such as water pH, ORP (Oxidation Reduction Potential), conductivity and hardness, $P_e$ is performance of the overall plant - e.g. if a water distribution system supplies 10\,million gallons per day, attacker intent may be to reduce it to 5-million gallons per day. The attacks are on 1\_LT\_001, 2\_LT\_002, 1\_MV\_002, 2\_MV\_003, and 1\_MV\_001 which form the $C_m$ dimension of the attack domain. For the dimensions considered in this paper, refer to Table\,\ref{table:attack_summary}. The attacks also affect the flow of water that falls along the $P_e$ dimension. $P_r$ is an empty set as the attacks do not affect the property dimension. Based on the above description, six attacks were designed and launched one at a time (refer to the Table\,\ref{table:attack_summary} for summary of all attacks). 

As we discussed in the attacker model, we derived the attacks from an intent of the attack. Based on the existing realistic attacks and incidents reported in the literature on water distribution systems, we considered the following intents in our experiments: 1)\,stop water supply to consumers, 2)\,damage water pumps in water distribution system, 3)\,overflow the water tanks, 4)\,wastage of water by leaking the pipe,  5)\,burst the water pipes, 6)manipulate the dosing mechanisms in a water distribution systems.  

 One might attempt to realize only one or more than one intent   (mentioned in Table\,\ref{table:attack_summary}) at a time. There are a couple of steps in going through to realize an intent: 1)\,understand the physical process, 2)\,based on the intent, identify the set of sensors or actuators to manipulate, and 3)\, control process to reach the intent. 
Initially, we understand the WADI process behavior and identify the set of sensors and actuators to be attacked in order to reach the intent. We divided the attacks into two categories based on the number of sensors and actuators attacked. A single-point attack is when only one sensor or actuator is attacked.  When the attack occurs on more than one sensor or actuator, it is classified as a multi-point attack. In Table\,\ref{table:attack_summary},  four single point  and two multi point attacks are listed. 

\begin{table}[tbh]
	\centering
    \small
	\caption{Summary of  attacks launched on WADI}
    \vspace{0.5 cm}
	\label{table:attack_summary}
	\begin{tabular}{|p{0.4 in}|p{0.7 in}|p{1.5 in}|p{0.7 in}|p{0.7 in}|}
		\hline 
\textbf{Attack No} & \textbf{Attack Sensor/Actuator} & \textbf{Intent} & \textbf{Start state($S_o$)}  & \textbf{End state($S_e$)}      \\ \hline 
        \multicolumn{5}{c}{\textbf{Single Point Attacks}} \\ \hline 
		1   & LIT  - 1\_LT\_001   & Block flow of water to ER tank   & 48\%  &  40\% \\ \hline
		2   & LIT - 2\_LT\_002 &  Stop flow of water to consumers and damage pump  &    & 80\%  \\ \hline
		3   & MV - 1\_MV\_002   & No flow of water to the consumers    &  Open & Close  \\ \hline
		4   & MV - 1\_MV\_001 & Block flow of water to raw water tank  &  Open  & Close \\ \hline

        
        \multicolumn{5}{c}{\textbf{Multi Point Attacks}} \\ \hline 
        
        5   &  1\_AIT\_002, 2\_MV003 & Supply contaminated water to the elevator tank  & 1\_AIT\_002 is 0.5 and 2\_MV003 is Close    &  1\_AIT\_002 is 6 and 2\_MV003 is Open     \\ \hline
        6    &  2\_MCV\_101, 2\_MCV\_201  &  Intermittent supply to consumer tank  &   Both  Close  &  Open both valves at 50\% \\ \hline
        
    \end{tabular}
\end{table}

\subsection{Execution of attacks}
We used the NiSpliot (see Section~\ref{attacking}) to launch the attacks listed in Table\,\ref{table:attack_summary}. The remaining subsection offers details of  each attack.

\subsubsection{Attack\,1: Attack on 1\_LT\_001}
\label{sec:sharedVariables}

This is an attack on level indicator  1\_LT\_001. This level indicator measures the level in the raw water tank (stage 1). The related shared variable is stored at the path {\it P1-CompactRIO/HMI\_HOST/HMI\_1\_LT\_001} and contains measurements for the water level in raw water tank\,1. The shared variable cluster can be broken further into the following variables.

\begin{itemize}
	\item PV - Process value measures water level.
	\item SIM PV - Process value  used in simulation Mode.
	\item SIMULATION - This variable is a boolean, sets whether the PV is to be used in the simulation PV or the actual PV.
	\item SAHH - Set point Alarm High High, the \state{HH} alarm default is 90.
	\item SAH - Set point Alarm High, the High (\state{H}) alarm set point default is 70.
	\item SAL - Set point Alarm Low, the Low (\state{L}) alarm set point default is 60.
	\item SALL - Set point Alarm Low Lo (\state{LL}), the Low Low alarm set point default is 40.
	\item S EMPTY - Set point for the state in which the tank is considered empty, default is 35.
	\item A EMPTY - Alarm indicating S EMPTY is reached.
	\item AHH - Alarm indicating SAHH is reached
	\item AH - Alarm indicating SAH has been reached.
	\item AL - Alarm indicating SAL is reached.
	\item ALL - Alarm indicating SALL is reached.
\end{itemize}

In this attack the attacker sets SIMULATION to True and also sets Simulation PV to 40  while setting S\_EMPTY to 40 using a script written using the NiSploit library. Thus, the state of WADI moves from $S_o$=\{SIMULATION=False, S EMPTY=35, 2\_MV\_004=Open\} to $S_e$=\{SIMULATION=True, S EMPTY=40, 2\_MV\_004=Close\}.

\subsubsection{Attack 2: Attack on 2\_LT\_002}
\label{attack2}

This is an  attack on level indicator 2\_LT\_002. This level indicator measures ER tank-2 level in process\,2. The related shared variable is stored at the path P2-CompactRIO/HMI\_HOST/HMI\_2\_LT\_002 and contains measurements for the water level in ER tank-2. The shared variable cluster can be broken further into smaller variables  as described in Section\,\ref{sec:sharedVariables}. In this attack the attacker sets PV to 80 by running a continuous loop. The state of valves and pumps remains unchanged, i.e. open and running, but the level of  water falls in both the Raw Water Tank and the ER.

\subsubsection{Attack 3: Attack on Motorized Valve 1\_MV\_002}

This attack is on motorized valve  1\_MV\_002. This motorized valve is an actuator in process\,1, the related shared variable is stored at the path P1-CompactRIO/HMI\_HOST/HMI\_1\_MV\_002 and contains the current status of the respective motorized valve governing the flow of water to the drain.

The shared variable cluster can be broken further into smaller variables. The state of the system moves from $S_o$=\{1\_MV\_002=Close, 2\_MV\_004=Open\} to $S_e$=\{1\_MV\_002=Open, 2\_MV\_004=Close\}.
\begin{figure*}[htbp]
\centering

\begin{minipage}{.5\textwidth}
  \centering
  \includegraphics[width=\linewidth]{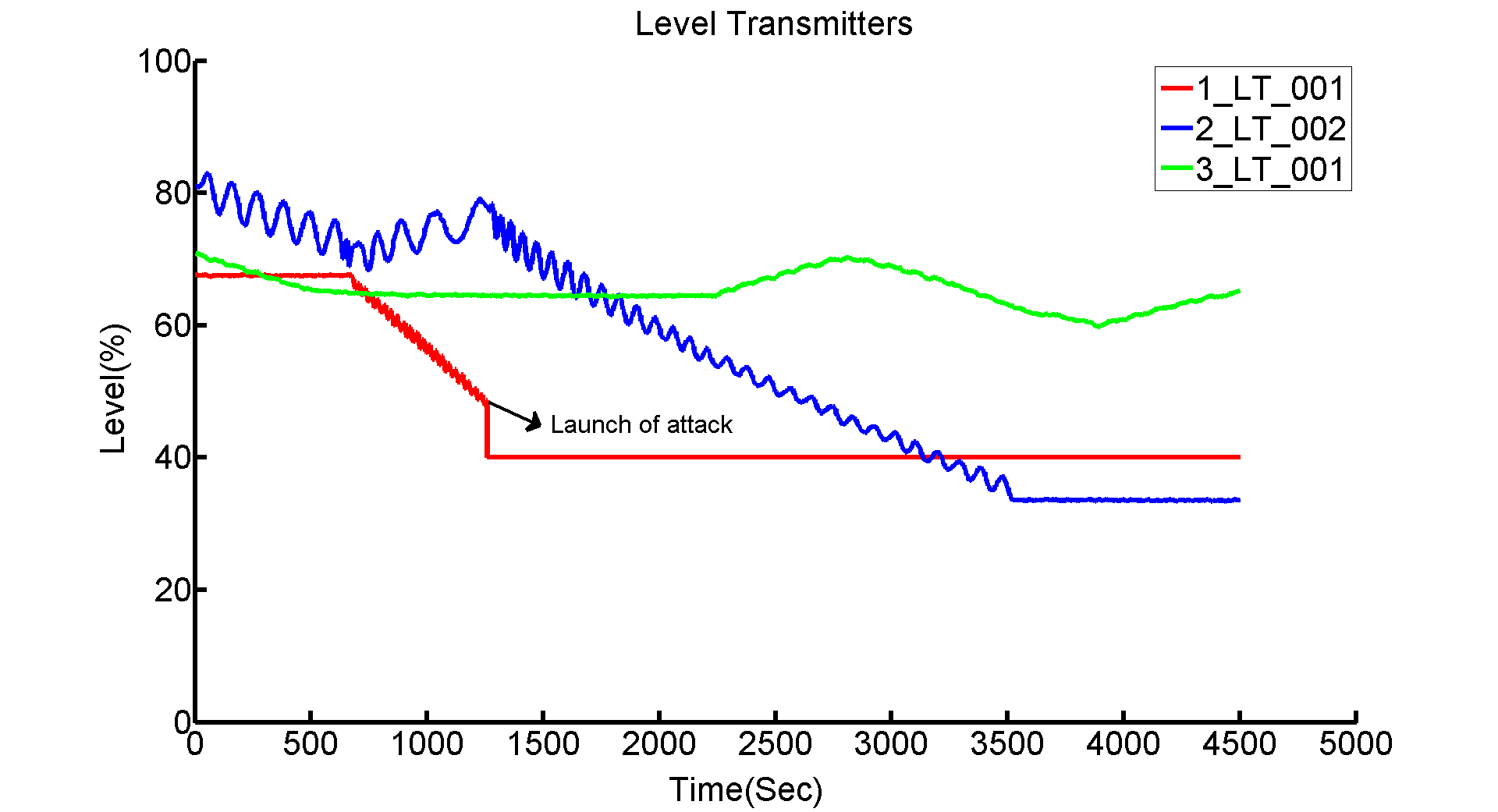}
 
	\caption[Attack 1: Attack on raw water tank level]{\small Attack 1: Water level readings of three stages.\\ Attacker brings the level of 1\_LT\_001 to 40\%. \normalsize}
	\label{fig:attacck1_tank}
\end{minipage}%
\begin{minipage}{.5\textwidth}
  \centering
  \includegraphics[width=\linewidth]{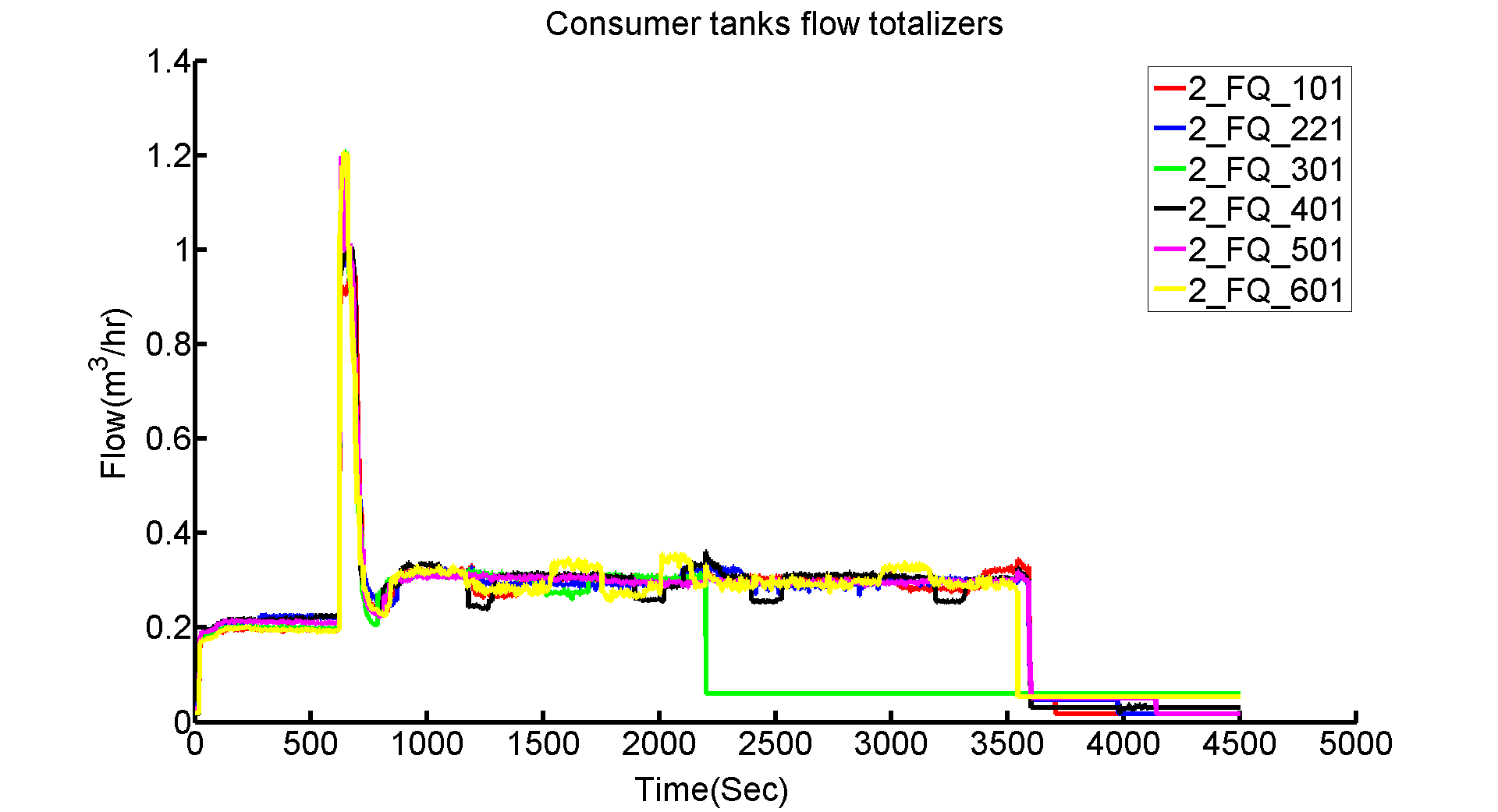}
	\caption{\small Attack 1: Flow to the consumer tanks and consumers are cut-off from water supply from little over 3500 seconds onwards. \normalsize}
	\label{fig:attacck1_totalizers}
\end{minipage}%
\end{figure*}

\begin{itemize}
	\item Auto - If set to True, the motorized valve works according to the programmed logic.
	\item Open Command -  open the valve
	\item Close Command - close the valve
	\item Reset - reset valve state to default state
	\item Available - Check if the Valve is available for control.
	\item Fully Open - Boolean indicating whether the Valve is fully open.
	\item Fully Close - Boolean indicating whether the valve is fully closed.
	\item Failed to Open - When the open command is sent but the valve could not be opened.
	\item Failed to Close - When the close command is sent and the valve could not be  be closed.
	\item Status - The current status of the valve.
	\item State - The current state of the valve, i.e. open or closed.
\end{itemize}

\noindent The attacker sets Auto to False and force opens the drain valve.

\subsubsection{Attack 4: Attack on Motorized Valve 1\_MV\_001}

This  attack is  on motorized valve  1\_MV\_001. This motorized valve is an actuator in process\,1. The related shared variable is stored at the path P1-CompactRIO/HMI\_HOST/HMI\_1\_MV\_001 and contains the current status of the motorized valve governing the inflow of water to raw water tanks.  The attacker sets Auto to False and sends the Close command. The state of WADI moves from $S_o$=\{1\_MV\_001=Open, 2\_MV\_004=Open\} to $S_e$=\{1\_MV\_001=Close, 2\_MV\_004=Close\}.

\skipnoindent In the previous sections, we described the single point of attacks. It is also possible an attacker can target multi points at a time, within the single stage and/or across multiple stages. However, in this study we investigated attacks on maximum two points. As shown in the Table\,\ref{table:attack_summary}, four two point attacks are launched on the system. 

In attack\,5, the attacker intention is to supply contaminated water to the elevator tank. In order to realize this intent attacker targets multistage multi point attack across the processes P1 and P2.  In this attack, attacker targets 1\_AIT\_002 in process1 and 2\_MV002 in process2. In attack\,6, the attacker intention is to cause intermittent supply to consumer tank. This is an single stage multi point attack, where attacker targeted two actuators (2\_MCV\_101, 2\_MCV\_201 ) in process P2. Initial and final states of the system during attack\,5 and attack\,6 are mentioned in Table\,\ref{table:attack_summary}.

\subsection{Results}
\label{sec:results}


The results show how an attacker is able to reach his intent. This kind of study is helpful to perform the impact analysis of the system. The remaining subsection presents the results for the attacks designed in the Table\,\ref{table:attack_summary}.

\subsubsection{Attack 1: Attack on 1\_LT\_001}
\label{subsubsec:attack1}



From Figure\,\ref{fig:attacck1_tank} it can be seen that the attack begins slightly after 1000~seconds when the  1\_LT\_001  is set to simulation mode with SIM PV at 40.  Figure\,\ref{fig:attacck1_tank} shows the attack on 1\_LT\_001 in which the attacker alters  the reading from 48\% to 40\% of the RW tank level which corresponds to a LowLow (\state{LL}) state. Since the raw water tank is in \state{LL} state the controller sends a command to open the PUB inlet valve, or the return water grid pump,  to fill the tank. Further, due to \state{LL} state of the RW tank there is no  flow of water from primary  to the secondary grid. It is to be noted that at the time of attack launch  on RW tank, the secondary grid is at 50\% of the maximum tank level. Therefore, the secondary grid supplies water to the consumer tanks until it reaches to 35\% of the maximum tank level which is considered  an ``Empty" state. The secondary grid tank level (2\_LT\_002)   behavior is  shown in  Figure\,\ref{fig:attacck1_tank}. Figure\,\ref{fig:attacck1_totalizers} indicates that no water flows to the consumers when the secondary grid tank is in  Empty state. Further, the RW tank overflows as there is no flow from the  primary grid to the secondary grid though there is continuous supply of water to RW tank through the PUB valve.

\begin{figure}[htbp]
  \centering
  \includegraphics[width=0.8\linewidth]{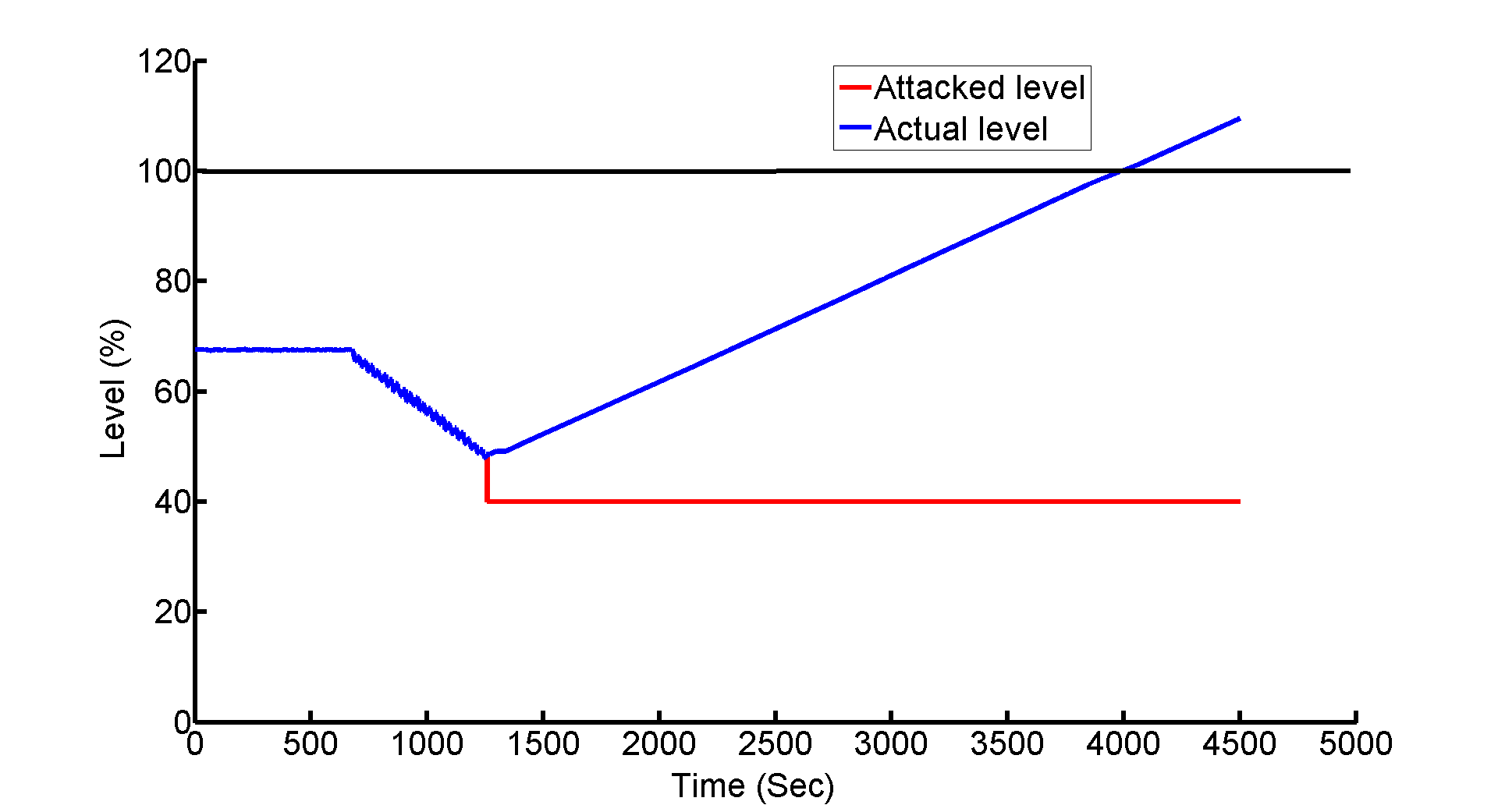}
	\caption[Attack 1: Attack on raw water tank level]{\small Attack 1: Actual level of the RW tank as it overflows. \normalsize}
	\label{fig:attack1_estimatedlevel}
\end{figure}%

\noindent It is possible to estimate from first principles the  water level in a tank. Mass balance equations, in continuous and discrete forms, for the change in water level  $h$ for a given input $Q_{in}$ and output $Q_{out}$, flow rate, as  follows,

\begin{eqnarray}
A\frac{dh}{dt}=Q_{in}-Q_{out},\\
h(t+1)=h(t)+\frac{\Delta t (Q_{in}(t)-Q_{out}(t))}{A}\label{eq:discrete},
\end{eqnarray}

\noindent where $A$ is the cross sectional area of the tank. Assuming linear dynamics, $Q_{in}$ and $Q_{out}$ are either 0 (when valve closes) or constant (when valve opens). We use Eq\,\ref{eq:discrete}  to estimate the tank level when a sensor is under attack. In this attack, the attacker sets the value of 1\_LT\_001 to 40\% which corresponds to \state{LL} state. Consequently the outlet flow rate  $Q_{out}$ is zero. Hence, Eq\,\ref{eq:discrete}  reduces to the following
\begin{equation}
h(t+1)=h(t)+\frac{\Delta t (Q_{in}(t))}{A}.\label{eq:reducedEquation}
\end{equation}
Using Eq.\,\ref{eq:reducedEquation}  we estimate the actual level  of the tank. As  in  Figure\,\ref{fig:attack1_estimatedlevel} the tank overflows when the attacker sets a constant value to 40\%.   


\subsubsection{Attack 2: Attack on 2\_LT\_002}

\begin{figure}[htbp]
	\centering
	\includegraphics[scale=0.33]{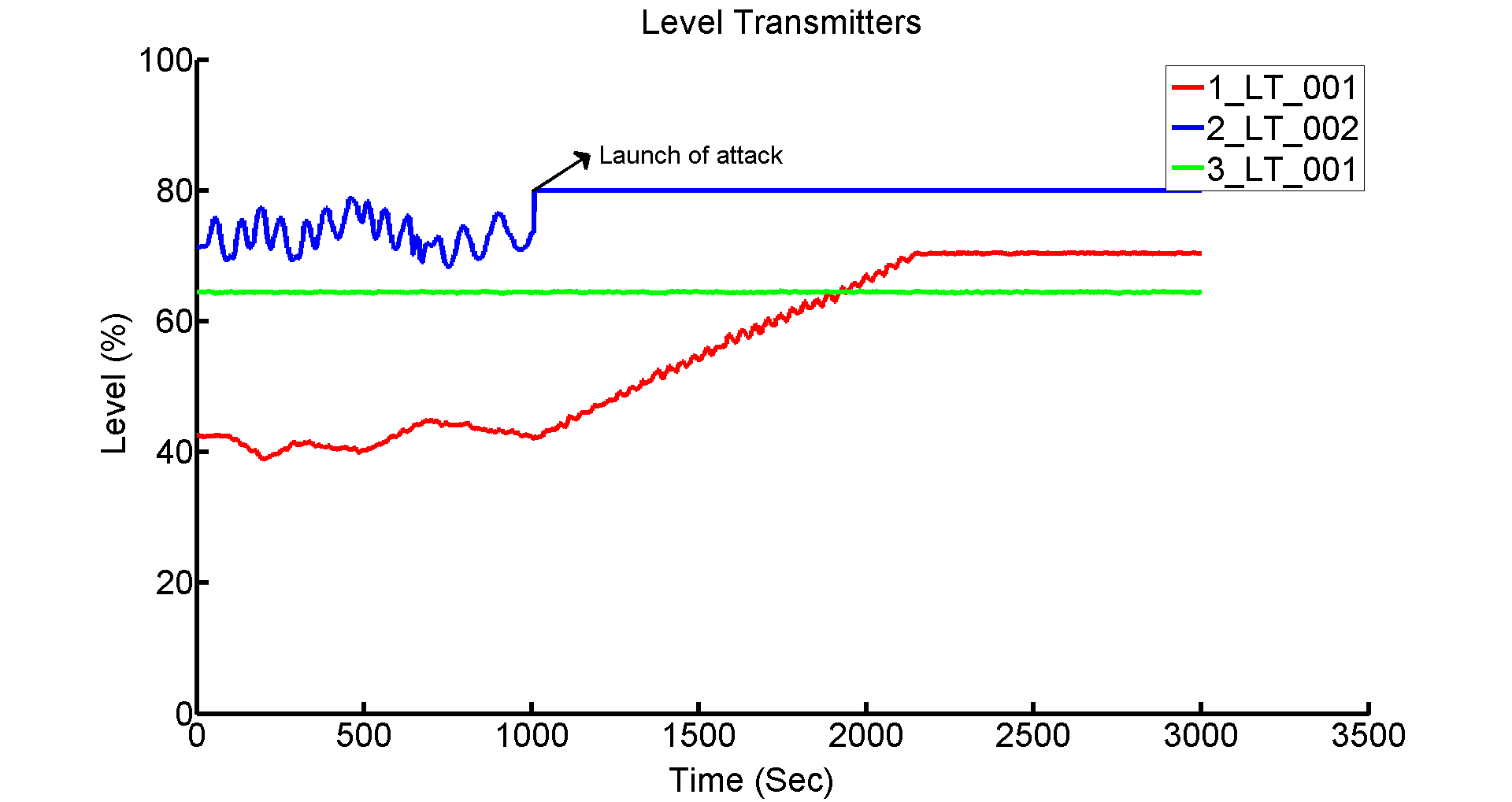}
	\caption[Attack 2: Water Level]{\small Attack 2: Water level readings of tanks. Figure shows launch of attack on 2\_LT\_002 at approximately  $1000$ seconds. \normalsize}
	\label{fig:attack2_level}
\end{figure}

\begin{figure}[htbp]
\centering
\includegraphics[scale=0.33]{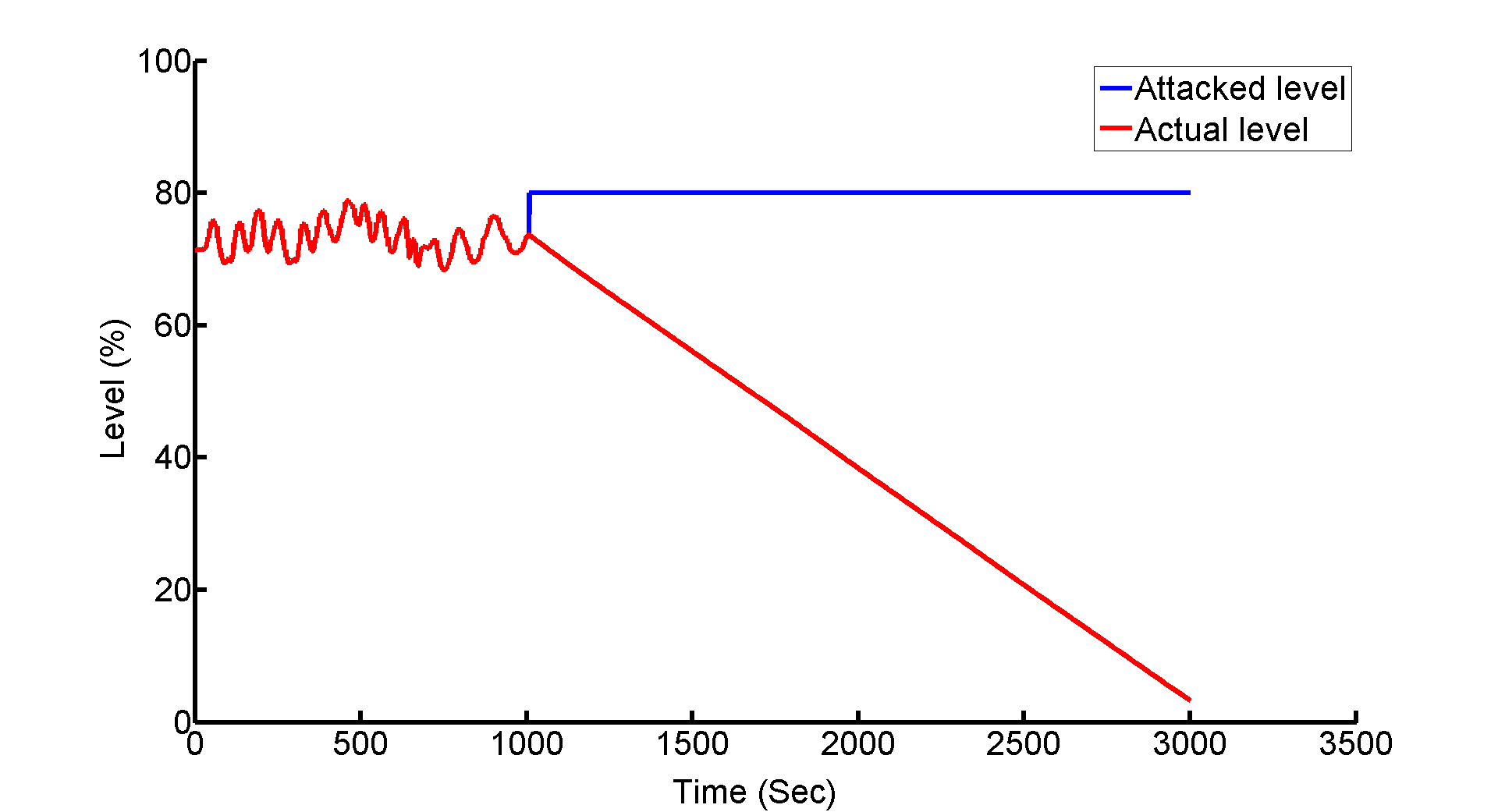}
\caption{\small Attack 2: Actual water level of ER tank (2\_LT\_002) goes into Empty state. \normalsize}
	\label{fig:attack2_level_actual}
\end{figure}

In Figure\,\ref{fig:attack2_level} it can be seen that the attack begins after  1000\,seconds when 2\_LT\_002  is set to 80\% of the tank level which corresponds to  High (\state{H}) state. This leads to no flow of water from the RW tank to ER tank. However, the ER tank continuously supplies water to the consumers.  After sufficient time has elapsed,   the actual  ER tank level moves to Empty state as  seen in Figure\,\ref{fig:attack2_level_actual}. It can be observed that in this situation the booster pump will be running continuously assuming that ER level  is at \state{H}. Consequently the booster pump will run dry and may be damaged unless a physical protection, e.g., a temperature cut off, are installed. Further, supply to the consumers stops completely.

\subsubsection{Attack 3: Attack on Motorized Valve 1\_MV\_002}
\begin{figure}[htbp]
	\centering
	\includegraphics[scale=0.33]{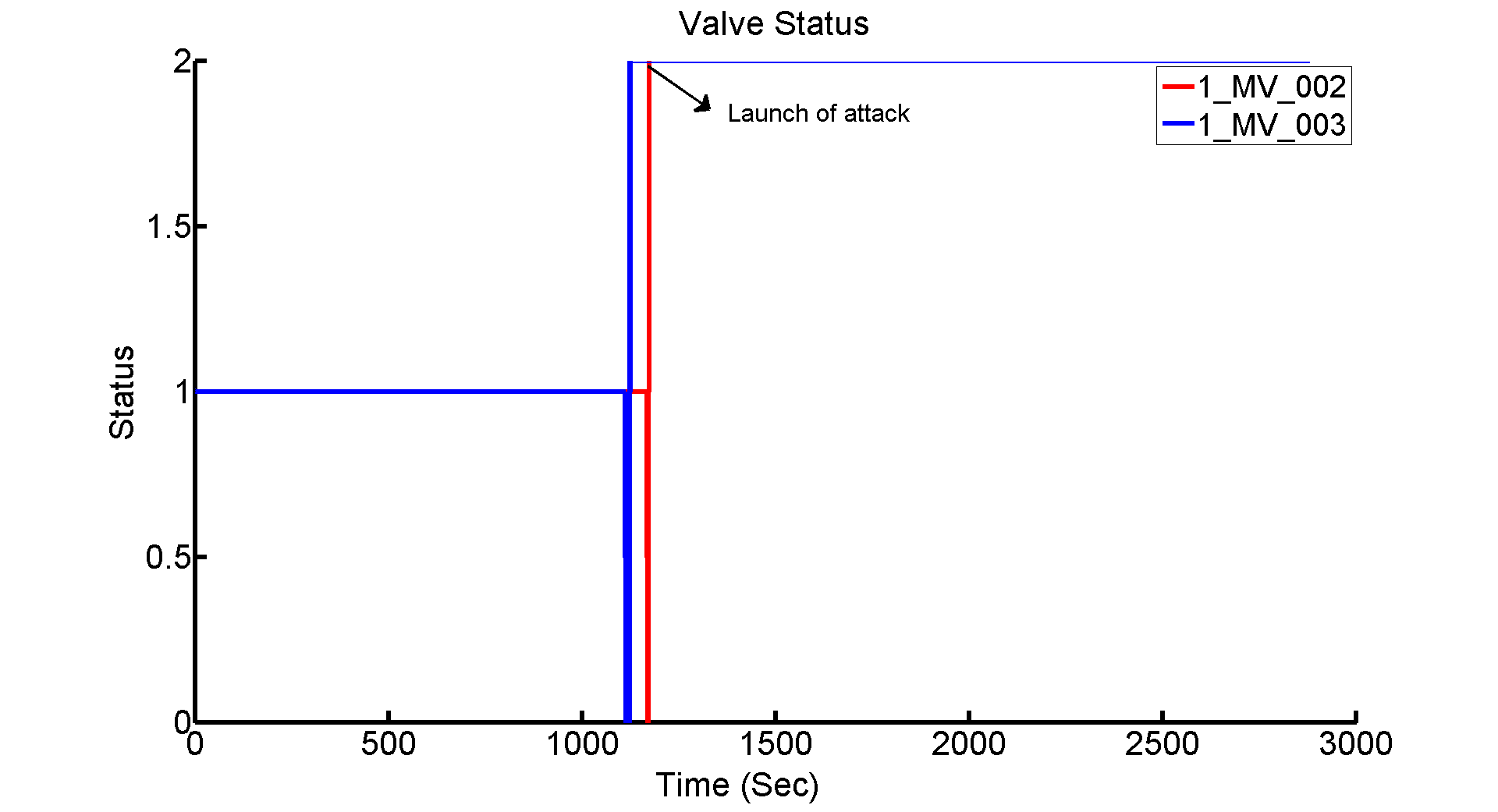}
\caption{ \small Attack 3: Attack on valves 1\_MV\_002 and 1\_MV\_003. \normalsize}
	\label{fig:attack3_valve}
\end{figure}

\begin{figure}[htbp]
	\centering
	\includegraphics[scale=0.33]{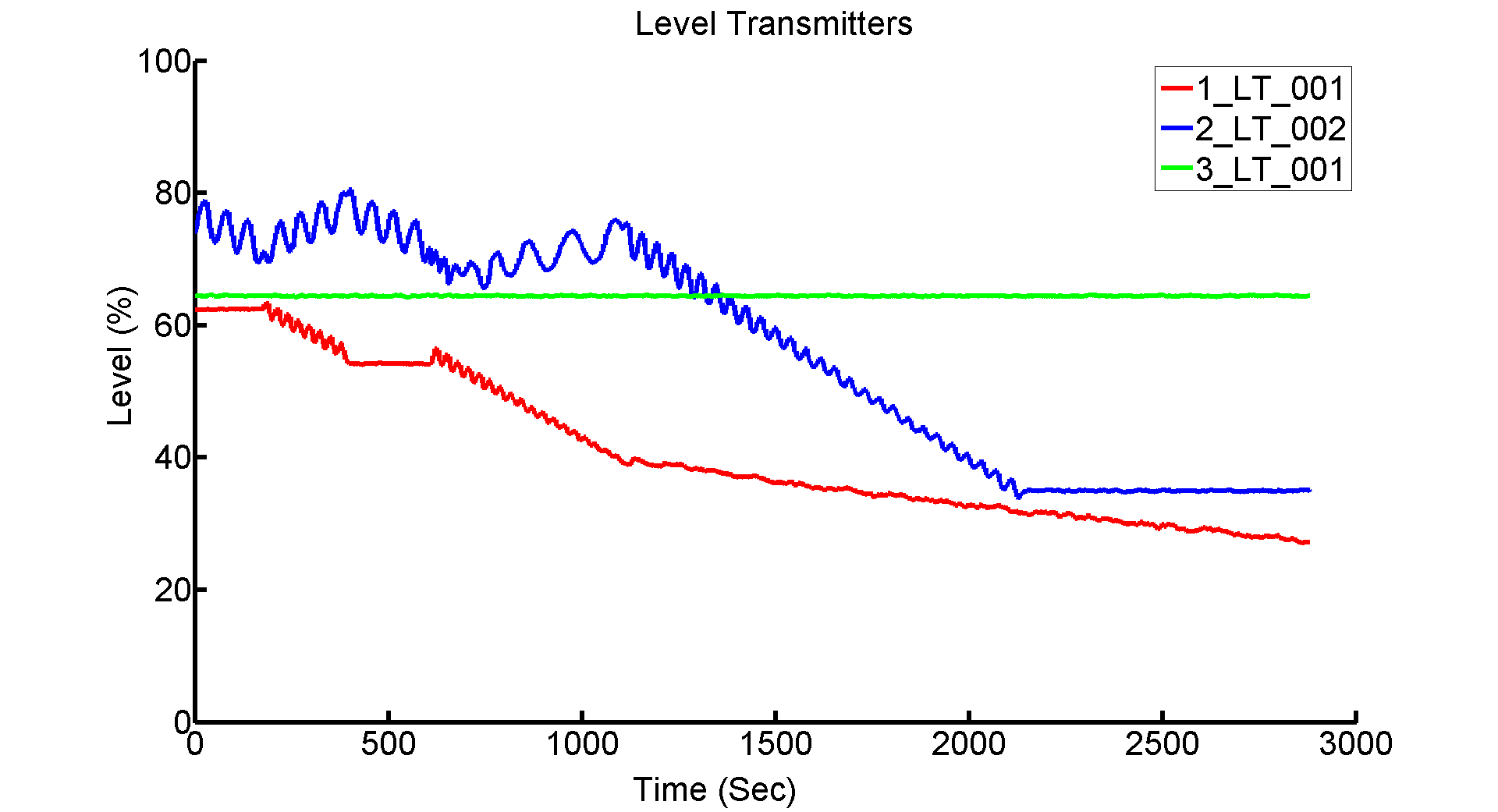}\
\caption{\small Attack 3: Water tank levels of 1\_LT\_001 reduces gradually. At $\approx2250$ s  2\_LT\_002 reaches to Empty state (35\% of tank level). \normalsize  }
\label{fig:attack3_level}
\end{figure}

\begin{figure}[htbp]
	\centering
	\includegraphics[scale=0.33]{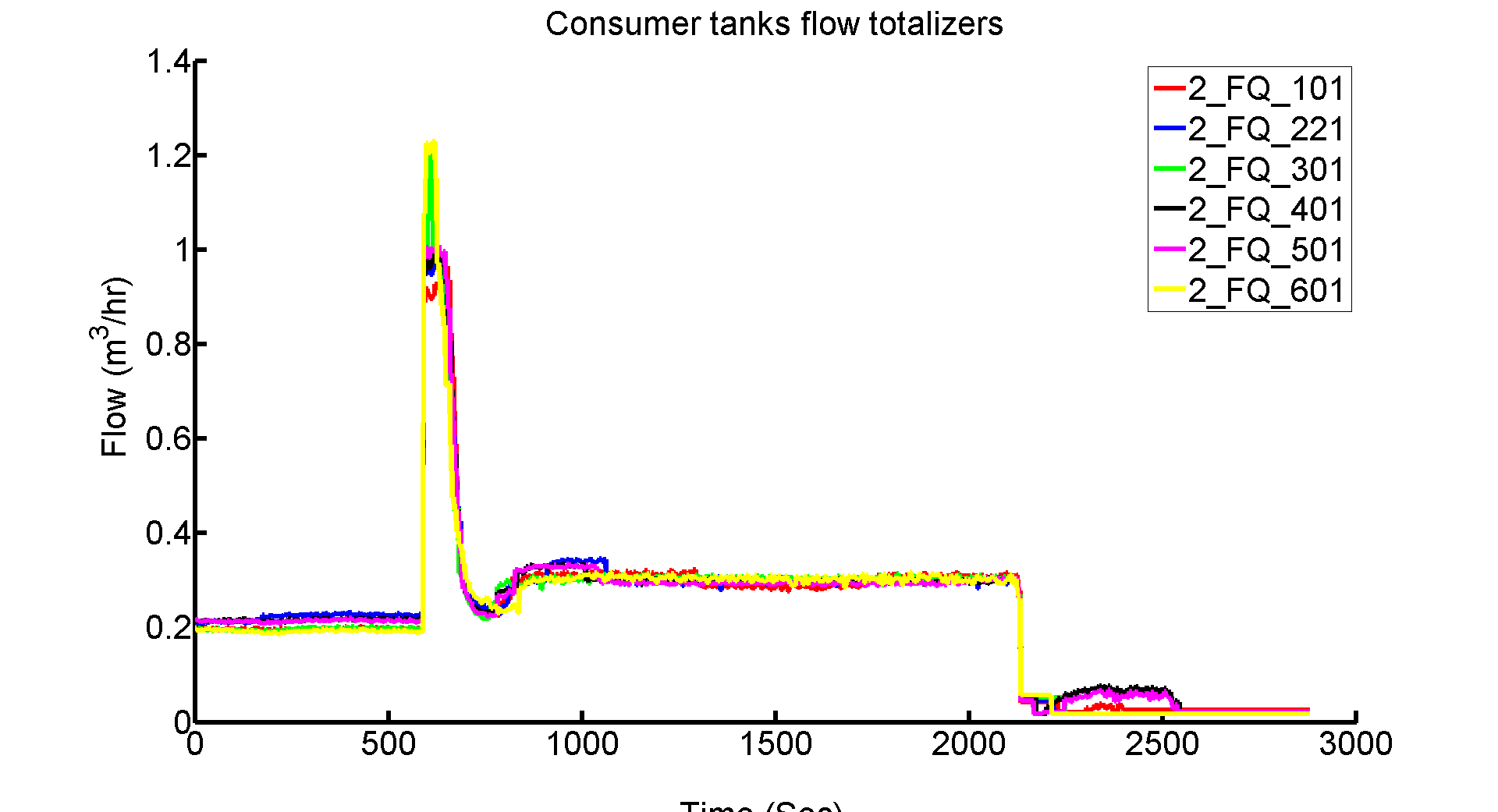}
	\caption[Attack 3: Water Flow]{\small Attack 3: Water flow to the consumers. \normalsize}
	\label{fig:attack3_flow}
\end{figure}


In Figure\,\ref{fig:attack3_valve} it can be seen that the attack begins after   1000~seconds when  valves 1\_MV\_002 and 1\_MV\_003   (also called drain valves) are  forced open. When these valves are open, water starts draining from the RW tank. Also, water is supplied to the ER tank when its level reaches the  \state{L} state. After some time water level in the RW tank reaches to  \state{LL} state and consequently PUB inlet valve, or return water grid pump, turns on to fill the tank. Note that water filling (through the PUB valve or return water grid) and draining (through 1\_MV\_002 and 1\_MV\_003) happens simultaneously. This leads to the water level in the tank at 40\% or below depending on the inlet and drain water flow rate. Figure\,\ref{fig:attack3_level} shows that water level falls below 40\% gradually leading to  no water supply from RW tank to the ER tank. Consequently water supply will be stopped to the consumer tanks (shown in Figure\,\ref{fig:attack3_flow}) when the level in the ER tank falls to 35\% of the maximum tank level.

\subsubsection{Attack 4: Attack on Motorized Valve 1\_MV\_001}
\begin{figure}[htbp]
	\centering
	\includegraphics[scale=0.33]{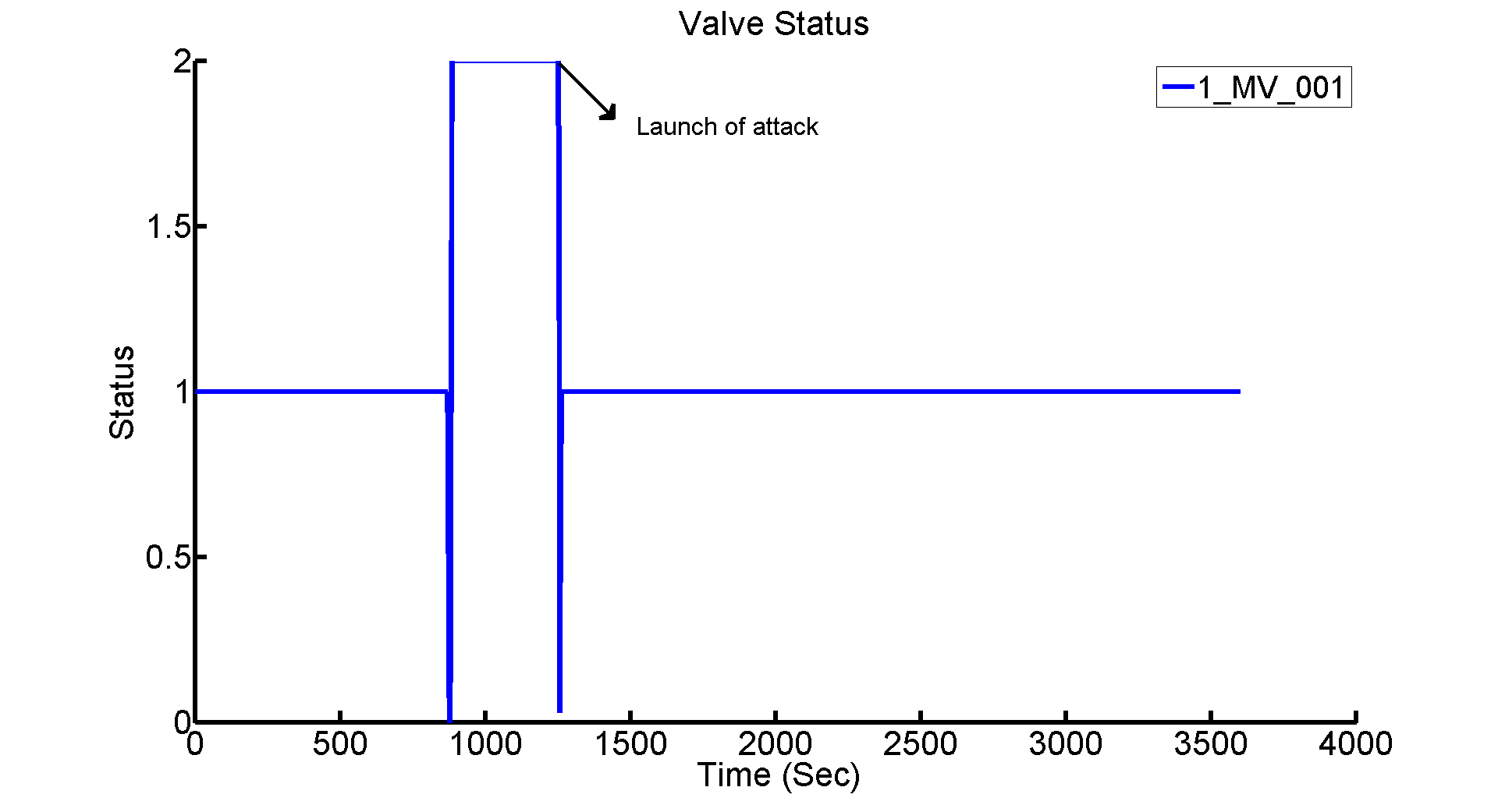}
	\caption{\small Attack 4: Attack on valve 1\_MV\_001 at approximately $1250$ seconds \normalsize}
	\label{fig:attack4_valve}
\end{figure}

\begin{figure}[htbp]
	\centering	\includegraphics[scale=0.33]{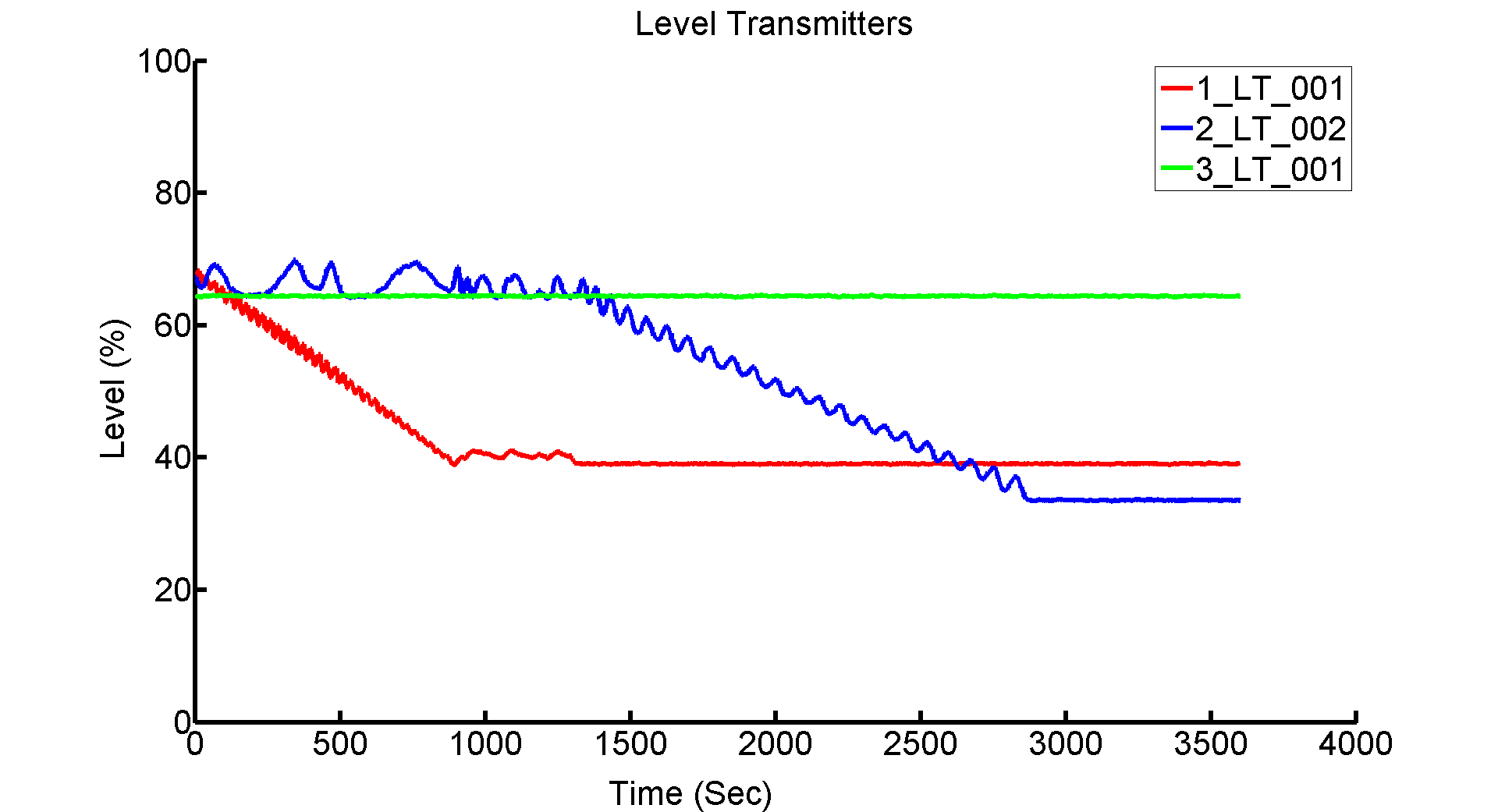}
	\caption{\small Attack 4: Water tank levels when 1\_MV\_001 is attacked. \normalsize}
	\label{fig:attack4_level}
\end{figure}
As in Figure\,\ref{fig:attack4_valve} the attack begins after 1000\,seconds when  1\_MV\_001 valve is forced shut. This leads to no water flow  into the RW tank. Figure\,\ref{fig:attack4_level} shows that RW tank level is kept  at 40\% as a result of the attack. Hence, there is no flow  from the RW to the ER tank. However, the ER  tank continuously supplies   water to the consumers. It can be observed from Figure\,\ref{fig:attack4_level} that ER tank level reaches Empty state after sometime  and there is no water flowing to the consumers.

\subsection{Multi point attacks}
\subsubsection{Attack 5}

\begin{figure}[htbp]
	\centering
	\includegraphics[width=\linewidth]{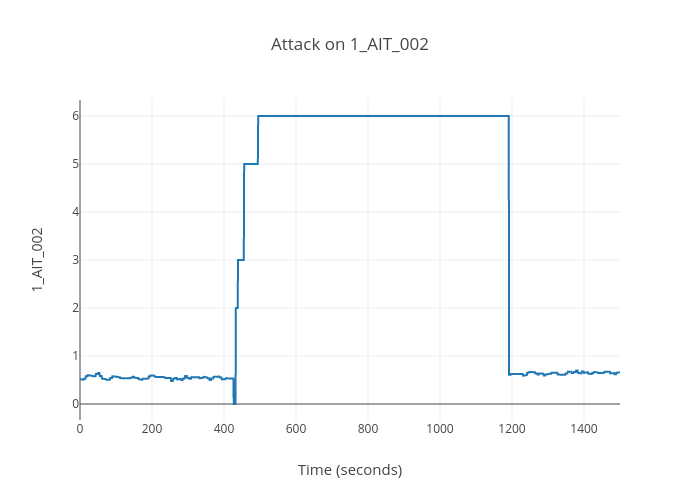}
	\caption{\small Attack 5: Attack on 1\_AIT\_001 \normalsize }
	\label{fig:attack9_AIT}
\end{figure}

\begin{figure}[htbp]
	\centering
	\includegraphics[width=\linewidth]{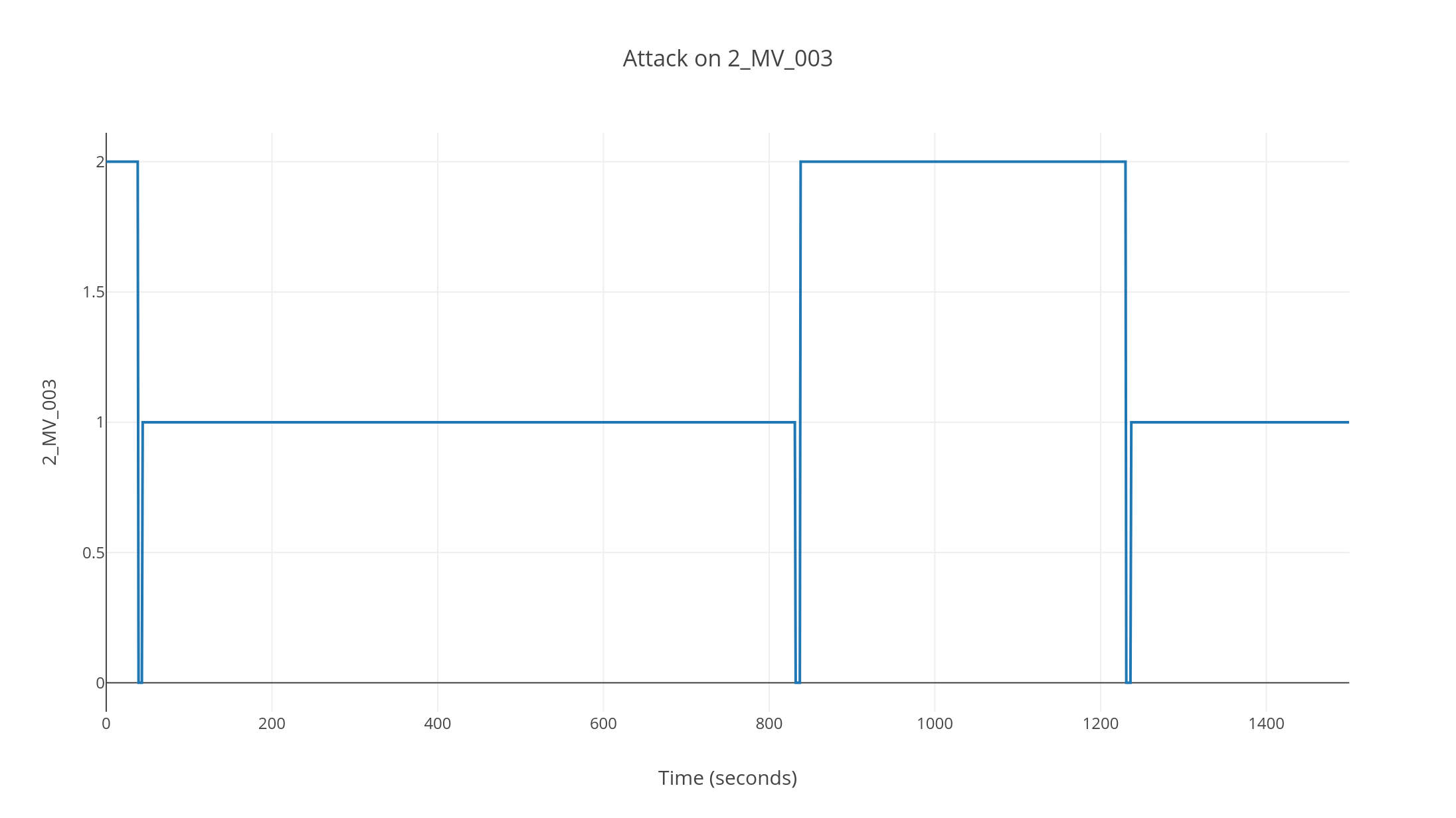}
	\caption{\small Attack 5: Attack on 2\_MV\_003 \normalsize}
	\label{fig:attack9_valve}
\end{figure}

In this attack, attacker launches multi point attack on 1\_AIT\_001 and 2\_MV\_003 as shown in Figure\,\ref{fig:attack9_AIT}\,and\,\ref{fig:attack9_valve} respectively. Initially, the attacker  manipulates 1\_AIT\_001 value from 0.5 to 6 which is above threshold at around 400 seconds. And, at around 500 seconds the attacker intentionally tries to open the inlet valve (2\_MV\_003) of elevated reservoir tank. As a result water from the raw water tank will be pumped to the elevated reservoir tank. Therefore, the attacker successfully achieves his goal by launching attack on 1\_AIT\_001 and 2\_MV\_003. 

\vspace{0.4 cm}

Similarly, attacks 6 is launched on the system to achieve his goals based on the attacker intentions. 
\section{Discussion}
\label{sec:discussion}



\noindent Next we summarize  what we learned during this investigation and provide answers to research questions stated earlier. 

\skipnoindent {\em Value of a testbed}: Researchers have studied\,\cite{aminLitricoSastryBayen2013PartI,taormina2017characterizing}  the attacks on water distribution systems. However, these studies have concentrated on small systems with a few  sensors and actuators, and thus are not adequate  
to investigate cyber attacks on larger systems. Characterization of cyber attacks on water distribution systems\,\cite{taormina2017characterizing} launched in a simulated environment may not be realistic though they do offer hints on the design of experiments reported here.  The study reported here overcomes the limitations of past studies by using a realistic water distribution system as the testbed, namely WADI.

\skipnoindent RQ1: {\em How do cyber attacks impact a water distribution system?}:

Section~\ref{sec:results} describes how six attacks affect the water distribution process in WADI. In summary, an attack may lead to any one or more of the following undesirable consequences: (a)~tank overflow, (b)~pressure drop at the consumer end, (c)~no water at consumer end, and (d)~equipment damage. In addition to the six attacks mentioned in Section~\ref{sec:results}, several other attacks can be launched on WADI. For example organic and inorganic contaminants may be added to water and the chemical sensors compromised\,\cite{palleti2016sensor} so that the attack is not detected. WADI also has a leakage simulator that can be used to launch leakage or water theft attacks. Such attacks and their impact on WADI will be study in the future.

\skipnoindent RQ2: {\em How does knowing the response of a CPS to one or more cyber attacks, help in designing an attack detection mechanism?}: 

Traditional attack detection is often based on network traffic monitoring.\,\cite{BaigAhmadSait} Proposed water marking schemes are based on control theory.\,\cite{kwon2013security}  It is well understood that cyber attacks or faults on the system affect specific  sensor readings. 

Future research will focus on the detection of  attacks such as those described  in Section~\ref{sec:attackDesign}. There exist several detection mechanisms in the literature. One such  mechanism is based on invariants derived from plant design.  A ``process invariant," or simply an invariant\,\cite{adepuMathurAsiaCCS2016} is a mathematical relationship among ``physical" and/or ``chemical" properties of the process controlled by the PLCs in a CPS. These invariants aid in  detecting such attacks. For example, attack~1 in Section~\ref{sec:attackDesign}  can be detected as follows. In this attack, attacker sets the raw water tank level to \state{LL} state and as a result 1\_MV\_001 opens to fill the tank.  Further, the tank level is not rising even though   the inlet valve is open and also there is no outflow from this tank. One can write the invariant for the valve and the tank level as follows. If the tank level is in \state{LL}  and the inlet valve opens, then after sufficient time the tank level should  rise to \state{L} or \state{H} state. However, in this case the tank level neither reaches \state{L}  nor the \state{H} state. Clearly, in this case the invariant is violated and hence the attack is detected. Therefore, these kinds of  invariants are useful in attack detection.  Note that violation of an invariant does not necessarily imply that there is a cyber attack; it could also be due to communication or component failure.

\section{Related Work}
\label{sec:relatedwork}


\skipnoindent \textit{Open research challenges:} 
Researchers have presented challenges in safety and security against cyber attacks that  need to be addressed while designing a CPS\,\cite{cardenasAminSastry,Lee,sabaliauskaiteAdepu2017}. 
Sajid et al.\,\cite{sajidAbbasSaleem} explained the integration of IoT and SCADA systems with a focus on security and how to integrate and  create  intelligent ICS using  the Internet. 
Humayed et al. \cite{humayedLinLiLuo}  surveyed literature on cyber physical systems security, and presented a orthogonal framework consists of security, components, system perspectives.  They focused mainly on four CPS systems such as  ICS, smart grids, medical devices, and smart cars. 

\skipnoindent \textit{Attack modeling and analysis:} 
Attacks have been modeled as noise in sensor data\,\cite{kwon2013security}. Attack models designed specifically for CPS include a variety of deception attacks including surge, bias, and geometric\,\cite{cardenasAminLinHuangHuangSastry}. Such models have been used in experiments to understand the effectiveness of statistical techniques in detecting cyber attacks.
The attacks designed in this work are based on a cyber-physical attacker model\,\cite{adepuMathurCompsac2016}. Jajodia et al. \cite{jajodiaNoel2010} proposed a detailed procedure for modeling cyber systems using attack graphs. Such graphs model practical vulnerabilities in distributed networked systems.  Chen et al.\,\cite{chenKalbarczykZbigniewNicolSandersTanTempleWilliamTippenhauerHoaYau} have proposed argument graphs as a means to capture the workflow in a CPS. The graphs are intended to assess a system in the presence of an attacker. The graphs are formed based on information in the workflow such as use case or state, physical system topology such as network type, and an attacker model such as an order to interrupt, power supply, physical tampering, network connection, denial of service, etc. Typed graphs\,\cite{bhaveKroghGarlanSchmerl} and Bayesian defense graphs \,\cite{sommestadEkstedtJohnson} are a few other important contributions to the modeling of cyber attacks. 


\skipnoindent \textit{Attacks on water systems:} The first well known attack on water supply was Maroochy Shire \cite{abrams2008malicious} in 2000 in Australia. Industrial Control Systems Cyber Response Team\,\cite{icsCERTAdvisory}  has reported several attacks on water systems and remedial actions to protect against these. Amin et al.\,\cite{aminLitricoSastryBayen2013PartI,aminLitricoSastryBayen2013PartII} studied attacks on water canal systems and presented attack detection methods based on control, hydrodynamic models. However, this paper focuses on an ICS system consisting of a few sensors and actuators. The formal approach\,\cite{eunsukAdepuJacksonMathur,patlolla2018approach} is used to analyse the security of a water treatment system. We aim at  investigating the impact of attacks on a larger system such as WADI,  which has more than 100 sensors and actuators. 
Riccardo et al.\,\cite{taormina2017characterizing} presented a modeling framework to characterize the cyber physical attacks on water distribution systems. This framework consists of a few categories of attacks and EPANET simulation models. The analysis is applied to C-Town network to show the usage of the framework. This work is mostly performed in a simulation environment while the study reported here was performed on an operational water distribution system\,\cite{mujeebPalletiMathur}. This research is helpful to understand the differences between simulation based attack investigation in water distribution systems, real time water distribution attacks. Section~\ref{sec:discussion}  addresses these differences and advantages of the approach used in the current work.

\skipnoindent{\em Attack detection in water systems:} 
Mitchel and Chen surveyed\,\cite{mitchell2014survey} intrusion detection techniques for CPS. They presented existing works based on a classification tree. They also presented the advantages and limitations of the techniques. 
The use of invariants for detecting attacks on CPS has been proposed and evaluated by several researchers such as in\,\cite{gamageMcMillanRoth,adepuMathurAsiaCCS2016,adepu2018distributed}. In this work it is claimed that the use of controlled invariant sets in detecting cyber attacks uses little information about the controller and hence is useful for a large range of control laws. Yuqi et. al.\,\cite{Chen-Poskitt-Sun18a} proposed an approach for learning physical invariants that combine machine learning with ideas from mutation testing. Data driven\,\cite{goh2017anomaly,linAdepu2018tabor} approaches for attack detection is studied on a water treatment system.  

\skipnoindent Security of cyber physical systems are also studied as decision games \,\cite{FreyRashidAnthonysamyPintoAlbuquerqueaqvi}. 
The BATADAL\,\cite{taormina2018battle} is a battle of the attack detection algorithms competition on water distribution symposioum. The goal of the battle was to compare the different detection methods to detect cyber physical attacks. The BATADAL was conducted on a C-Town network, a real-world, medium-sized water distribution system operated through Programmable Logic Controllers and a Supervisory Control And Data Acquisition (SCADA). Total seven different teams participated in the BATADAL and their effectiveness of was evaluated in terms of time-to-detection and classification accuracy. This emphasis of dealing with real-life infrastructure and equipment for training and research is also seen in the development of Capture the Flag style gamification of an ICS testbed platform\,\cite{antonioli2017gamifying,adepu2018assessing}. The activity described in this paper is not a conventional competitive hackathon; as a combination of jam and hackathon, it emphasises hacking CPS platforms as a means to integrate, demonstrate and explore lines of research.

\section{Conclusions and Future work}
\label{sec:conclusions}
This paper reports an investigation into the response of an operational water distribution plant to cyber attacks. The outcome of the investigation points to the importance of testbeds in understanding  stealthy  and a varied set of  attacks and practical issues in  operational water distribution plants. The case study  also indicates that an attacker will likely be able realize an intent when adequate  resources are available and the required  accessibility exists. The work presented in this paper is a step towards realizing a safe and secure critical infrastructure. 

\skipnoindent Future work includes understanding more stealthy attacks and the implementation of  a  prototype defence mechanism in WADI. We plan to implement some of the attack detection mechanism mentioned in the related work section and assess in a real time water distribution system. 

 
 \section*{ACKNOWLEDGMENT}
 This work was supported in part by the National Research Foundation (NRF), Prime Minister's Office, Singapore, under its National Cybersecurity R\&D Programme (Award No. NRF2014NCR-NCR001-40, NRF2015NCR-NCR003-001) and administered by the National Cybersecurity R\&D Directorate. The WADI testbed is built with the support from Ministry of Defense, Singapore and SUTD-MIT International Design Centre (IDC).

\bibliographystyle{ios1} 
\bibliography{Security.bib}



\section*{Appendix}

\begin{tabbing}
SCADA   \hskip0.1in\=\kill
{\bf Acronyms}:\\[0.1in]
AI\> Analog Input\\
AIT\> Analyzer Indicator and Transmitter\\
AO\>Analog Output\\
AP\> Access Point\\
ARP\>Address Resolution Protocol \\
CPS\>Cyber-Physical System\\
CUSUM\>cumulative sum\\
DAQ\>Data acquisition\\
DB\>Data Base\\
DCS\>Distributed Control System\\
DDoS\> Distributed Denial of Service\\
DI\>Digital Input\\
DLR\>Distributed Logic Router\\
DO\>Digital Output\\
DoS\> Denial of Service\\
FIT\>Flow Indicator and Transmitter\\
HMI\> Human Machine Interface\\
ICS\>Industrial Control System\\
LIT\>Level Indicator and Transmitter\\
LSTM\>Long short-term memory\\
MITM\>Man In The Middle\\
MV\>Motorized valve\\
NI-PSP\> National Instruments Publish Subscribe Protocol\\
PLC \>Programmable Logic Controller\\
RTU\> Remote Terminal Unit\\
RIO\>Remote Input Output\\
SCADA \>Supervisory Control and Data Acquisition\\
VI\>Virtual Instruments\\
WADI\>Water Distribution\\

\end{tabbing}

\end{document}